\documentclass[
 aip,
 amsmath,amssymb,
reprint,
]{revtex4-2}
\usepackage{lipsum}

\usepackage{graphicx}
\usepackage{dcolumn}
\usepackage{bm}

\usepackage[utf8]{inputenc}
\usepackage[T1]{fontenc}
\usepackage{mathptmx}
\usepackage{etoolbox}
\usepackage{hyperref}

\makeatletter
\def\@email#1#2{%
 \endgroup
 \patchcmd{\titleblock@produce}
  {\frontmatter@RRAPformat}
  {\frontmatter@RRAPformat{\produce@RRAP{*#1\href{mailto:#2}{#2}}}\frontmatter@RRAPformat}
  {}{}
}%
\makeatother

\newcommand{\SPINX}{\affiliation{Spin-X Institute, School of Physics and Optoelectronics, State Key Laboratory of Luminescent Materials and Devices, Guangdong-Hong Kong-Macao Joint Laboratory of Optoelectronic and Magnetic Functional Materials, South China University of Technology, Guangzhou 511442, China}}
  
\newcommand{\IACS}{\affiliation{School of Physical Sciences, Indian Association for the Cultivation of Science, Jadavpur, Kolkata 700032, India}}

\begin{document}


\title{Critical scaling behavior in skyrmion host ferromagnet CrTe$_{1.38}$}

\author{Suman Kalyan Pradhan}
\altaffiliation{These authors contributed equally to this work.}
\email{suman1kalyan@scut.edu.cn}
\SPINX

\author{Tuhin Debnath}
\altaffiliation{These authors contributed equally to this work.}
\IACS

\author{Rui Wu}
\email{ruiwu001@scut.edu.cn}
\SPINX

\date{\today}

\begin{abstract}
Materials hosting diverse topological spin textures hold significant potential for spintronic applications. In this context, CrTe$_{1.38}$, a quasi-two-dimensional material, stands out due to its stable Néel-type skyrmion phase over a wide temperature range, both with and without an applied magnetic field 
[APL 125, 152402 (2024)]. Thus, it is a promising candidate for investigating complex magnetic phenomena, offering valuable insights into the underlying magnetic interactions. This study investigates the critical behavior of CrTe$_{1.38}$ near $T_\text{C}$ by measuring DC magnetic isotherms. A systematic analysis of these isotherms with the magnetic field applied along the easy axis allows us to determine the asymptotic critical exponents: $\beta = 0.314$, $\gamma = 1.069$, and $\delta = 4.556$, where the Widom scaling law and scaling equations are verified the self-consistency and reliability. In this system, the magnetic exchange coupling $J$($r$) is the long-range type and decays spatially at a rate slower than approximately $\approx r^{-4.651}$. Most notably, a series of vertical lines in the low-field region of the initial magnetization curves below $T_\text{C}$ supports the existence of a skyrmion phase in this compound. 
\end{abstract}

\maketitle

\textbf{Keywords:} 
Néel-type skyrmion; 
Critical scaling; 
{\section{Introduction}}

Universality is a fundamental concept in condensed matter physics, characterizing the general scaling behavior and symmetries of thermodynamic observables, near a critical point \cite{Wilson}. As the system approaches the critical point, the magnitude, and the frequency of fluctuations in the relevant order parameter typically intensify. The second-order transition from the paramagnetic (PM) to the ferromagnetic (FM) phase is a classic example where these concepts have been thoroughly investigated \cite{PhysRevB.91.024403,Liu2016,Lin,Liu}.

The successful mechanical exfoliation of monolayer graphene \cite{Novoselov2004,Geim2007} has sparked significant interest in two-dimensional (2D) materials \cite{Novoselov2005,Zhou2018} within the fields of materials science and condensed matter physics. This enthusiasm arises from the fundamental insights these materials offer \cite{HZeng2012}  along with their potential for scalable device applications \cite{Wang2012,Cheng2014,Bhimanapati2015}. For a long time, the study of 2D materials was primarily focused on understanding and optimizing their mechanical and optoelectronic properties, constrained by the limitations of the Mermin-Wagner theorem \cite{Mermin1966}. However, this focus shifted with the discovery of nontrivial quantum phenomena \cite{He2013, Xu2014, Gong2017}, which has paved the way for exciting possibilities for next-generation spintronic devices based on these 2D materials. Among the many synthesized 2D ferromagnets (FMs), a few exhibit truly intriguing physical properties in their pristine phase \cite{PhysRevX.11.031047,Bera}. Not only that critical analysis of several layered ferromagnets mimics a diverse range of theoretical model systems.  CrSiTe$_3$, for instance, exhibits characteristics of a 2D Ising system \cite{Liu2016}, while CrGeTe$_3$ belongs to the universality class of tricritical mean-field theory \cite{Lin}. In contrast, the critical exponents of CrI$_3$ lie between those of the 3D Ising model and the tricritical mean-field model \cite{Liu, Lin2018}. In the Fe$_3$GaTe$_2$ system, exponents belong to the 3D Heisenberg model and magnetic interaction is long-range type \cite{Algaidi2024}.

In this regard, the binary chromium (Cr)-based telluride family, Cr$_x$Te$_y$, has garnered significant attention due to its high Curie temperature, prominent magnetic anisotropy, and excellent stability in air \cite{Huang_2021, Zheng2023, Yao2023}. The characteristics of this family vary with the values of $x$ and $y$, making it a particularly notable system in this field \cite{Liu2018,Saha2022,Chi2023,Pradhan2024}. Recently, CrTe$_{1.38}$, a non-centrosymmetric compound within the Cr$_x$Te$_y$ family, with a Curie temperature of $T_\text{C}$ = 196 K, and significant coercivity, has emerged as a highly promising candidate. Its diverse range of topological spin textures makes it particularly noteworthy, as demonstrated by our group \cite{Pradhan2024}. Notably, the presence of stable Néel-type magnetic skyrmion bubbles within the temperature range of 100 K to 170 K, even in the absence of an external magnetic field, marks a key observation \cite{Pradhan2024}. 

Given the promising applications and rich physics of CrTe$_{1.38}$, a comprehensive investigation of its magnetic exchange is essential, not only for advancing fundamental physics, such as nonlinear magnetic ordering, but also for establishing a foundation for the future application of skyrmion states in spintronics. In this research work, the critical behavior is investigated through bulk DC magnetization measurements.
Using analytical techniques such as the modified Arrott plot and critical isotherm analysis, we determine the critical exponents $\beta$, $\gamma$, and $\delta$, which are self-consistent and reliable, verified by the Widom scaling law and the scaling equations. This critical behavior of CrTe$_{1.38}$ indicates long-range magnetic interactions. The results suggest that the magnetic interactions in this system are highly complex, as they do not conform to a single universality class.\\

\section{Experimental Details}
The CrTe$_{1.38}$ single crystals were grown using the chemical vapor transport (CVT) method. Details of the sample preparation and characterization can be found elsewhere \cite{Pradhan2024}. Magnetization measurements were conducted using a superconducting quantum interference device (SQUID, Quantum Design MPMS3), with a maximum applied magnetic field ($H$) of up to 60 kOe, over a temperature range from 400 K to 2 K. The crystal was mounted on a quartz tube, with the magnetic field applied both parallel to the ${ab}$ plane (in-plane, IP) and along the $c$ axis (out-of-plane, OOP). Lorentz–Fresnel magnetic imaging experiments were performed in magnetic field-free conditions (Lorentz mode) at 200 kV in a Thermo Fisher Talos F200 TEM. A liquid N$_2$ cooled TEM specimen holder (Gatan 636) was used to control the specimen
temperature. For transport measurements, a 4-terminal configuration with linear four-probe contacts was used. Silver (Ag) was deposited as the contact metal, ensuring excellent Ohmic contact over the entire temperature range.
\section{Results and Discussions}
\subsection{Preliminary characterization}

\begin{figure}
\centering
\includegraphics[width=1.0\columnwidth]{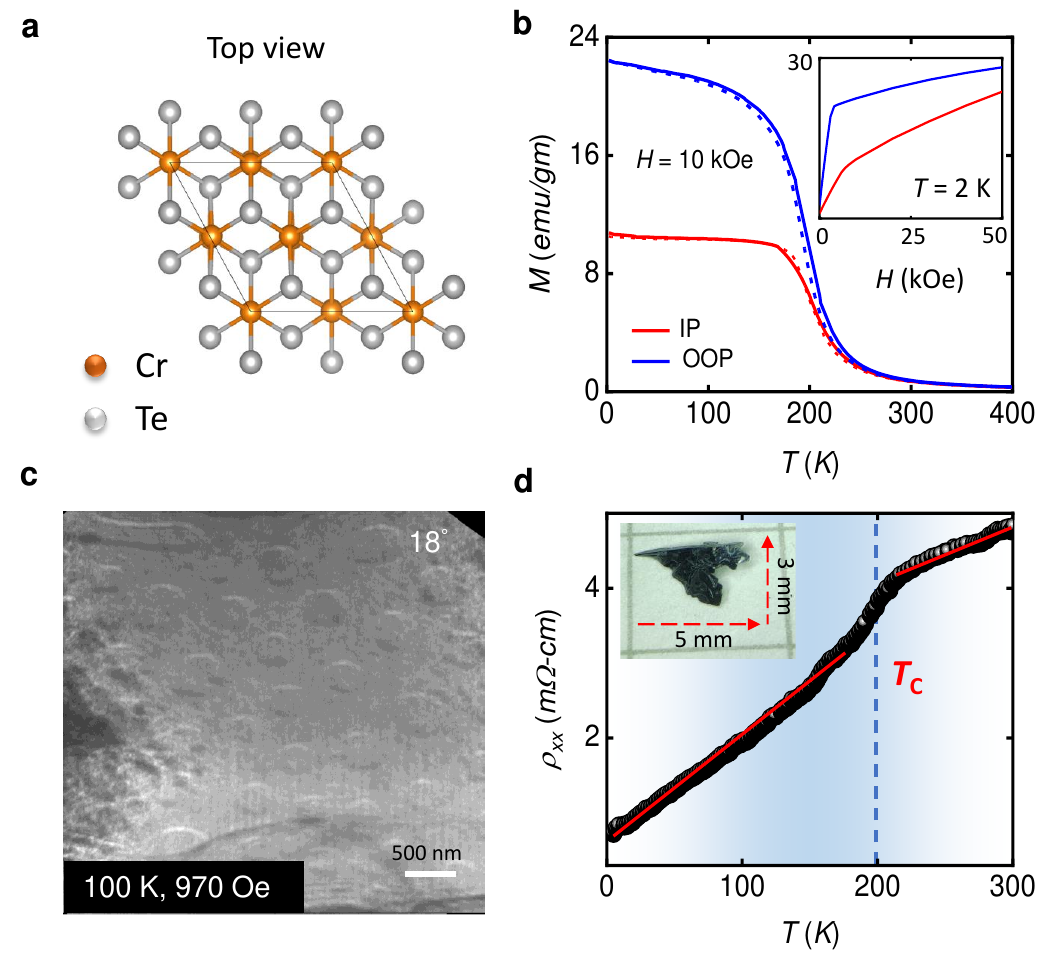}
\caption{\textbf{Preliminary  characterization :}  Schematic of the crystal structure, obtained from room-temperature powder x-ray diffraction (P-XRD) analysis, shown from the top view. The gray and golden yellow spheres represent the Cr and Te atoms respectively. The black box with the cross-sectional rectangle depicts the crystallographic unit cell. (b) Temperature dependence of magnetization measured in the external magnetic field $H$ = 10 kOe applied
along the ${ab}$ plane (IP) and $c$ axis (OOP) with zero-field-cooling (ZFC) and field-cooling (FC) modes. The ZFC and FC are represented by scattered and continuous lines. Inset displays the field dependence of magnetization measured at $T$ = 2 K along IP and OOP. (c) Over-focus Lorentz TEM observation of magnetic domain structures along the [001] direction at $T$ = 100 K and $H$ = 970 Oe. The sample thickness is 250 nm, with a tilt angle of 18$^{\circ}$ and a defocus distance of 1500 $\mu$m. The scale bar represents 500 nm. (d) Temperature-dependent zero field electric resistivity. The magnetic transition is well reflected around 200 K. Red solid lines guide eyes to observe the change of slope around $T_\text{C}$. The inset shows the photographic image of a grown crystal.}
\vspace{-0.45cm}
\label{charect}
\end{figure}

Fig.~\ref{charect}(a) illustrates the crystal structure of CrTe$_{1.38}$, which adopts trigonal crystal structure with non-centrosymmetric space group ${P3m1}$ (No.156) at room temperature \cite{Pradhan2024}. The structure reveals a total of 14 crystallographic sites, including eight Te sites and six Cr sites. The Cr atoms are centrally located, while the Te atoms arrange in a hexagonal pattern around them. The temperature-dependent magnetization, $M$($T$), measured with fields applied along the in-plane ($H$ $\parallel$ ${ab}$) and out-of-plane ($H$ $\parallel$ ${c}$) directions is shown in Figure~\ref{charect}(b). A clear paramagnetic (PM) to ferromagnetic (FM) transition is observed around 196 K ($T_\text{C}$). The initial magnetization curves $M(H)$ as displayed in the inset of Fig.~\ref{charect}(b) identifies the out-of-plane direction as the easy axis. Fig.~\ref{charect}(c) shows skyrmion bubbles at $T$ = 100 K and $H$ = 970 Oe. For instance, the spontaneous ferromagnetic ground state of CrTe$_{1.38}$ consists of Néel-type skyrmions in the temperature range of 100–170 K, even in the absence of an external magnetic field \cite{Pradhan2024}. Fig.~\ref{charect}(d) presents the in-plane electrical resistivity ($\rho_{xx}$) of CrTe$_{1.38}$ as a function of temperature, ranging from 5 K to 300 K, under zero magnetic field. The positive temperature coefficient of resistivity across the entire range indicates metallic behavior, with a relatively low residual resistivity ratio [RRR = $\frac{\rho(300 \text{K})}{\rho(5 \text{K})}$] of 6.6. No prominent peak corresponding to the paramagnetic-ferromagnetic (PM-FM) transition is observed; however, a change in slope is noticeable around $T = 196$ K. At low temperatures, up to 50 K, the resistivity follows a quadratic dependence, $\rho(T) = \rho_0 + aT^2$, consistent with Fermi liquid theory (not shown here). Here, $\rho_0$ represents the temperature-independent residual resistivity. The $T^2$ dependence at low temperatures suggests significant electron-electron scattering, a hallmark of itinerant ferromagnetic compounds.

\subsection{Critical Behavior}

\subsubsection{Paramagnetic to Ferromagnetic phase transition}

\begin{figure*}
\centering
\includegraphics[width=1.9\columnwidth]{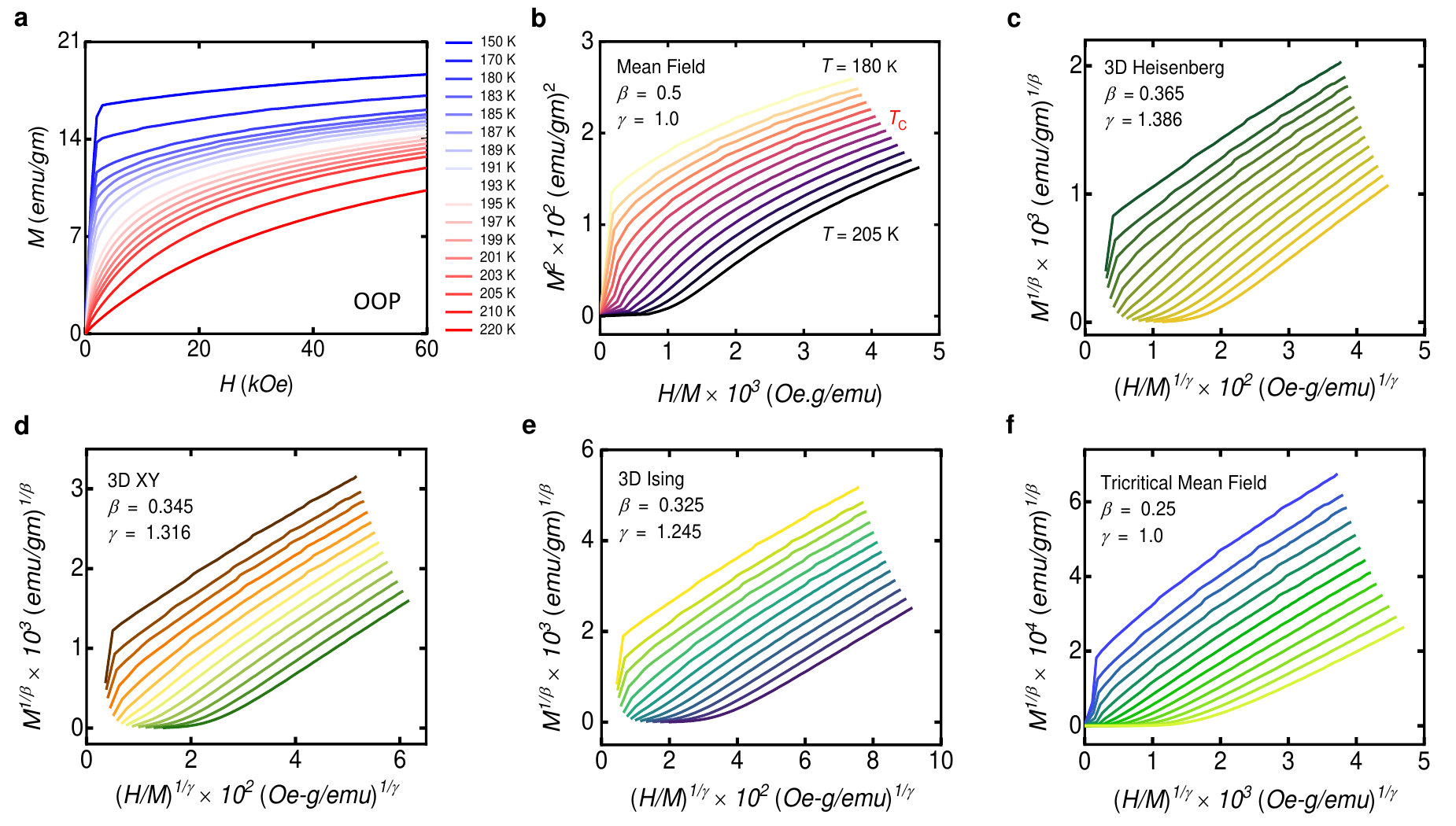}
\caption{\textbf{Magnetic isotherms and different theoritical models :} (a) Isothermal magnetization as a function of field $[M(H)]$ at the selected temperature along OOP. The modified Arrott plots of $M^{1/\beta}$ vs $(\frac{H}{M})^{1/\gamma}$ for $H//c$ with parameters of (b) mean-field model, (c) $3$D Heisenberg model, (d) $3$D XY model, (e) $3$D Ising model, (f) Tricritical Mean Field model.}
\vspace{-0.45cm}
\label{Fig 2-1}
\end{figure*}

To investigate the critical behavior of the paramagnetic (PM) to ferromagnetic (FM) phase transition in CrTe$_{1.38}$, magnetization isotherms were measured for out-of-plane directions in the temperature range 100 K $\leq$ $T$ $\leq$ 225 K. The data for the initial isotherms were collected with a temperature interval of $\Delta$$T$ = 2 K, in the close vicinity of the Curie temperature (183 K - 205 K) [see Fig \ref{Fig 2-1}(a)]. Now, the Landau theory of phase transitions proposes that the free energy $G$ can be expanded as a power series in the order parameter $M$ \cite{PhysRev.108.1394} in the vicinity of the phase transition. At equilibrium condition, this leads to  $M^{2}$=$\frac{1}{4b}$$\frac{H}{M} - \frac{a}{2b}$$\epsilon$,
where $a$ \& $b$ are temperature-independent coefficient, and $\epsilon = \frac{T-T_\text{C}}{T_\text{C}}$ is the reduced temperature. The $M^{2}$ vs $\frac{H}{M}$ relationship around $T_\text{C}$ constitute the Arrott plot. According to the Landau Mean-field model, the Arrott plot should consist of a series of straight lines parallel to each other in the high-field region \cite{Kaul1985}. The slope of the Arrott plot in this high-field region provides insight into the nature of the phase transition in a magnetic system, as determined by Banerjee’s criterion \cite{Banerjee1964}. A negative slope suggests a first-order transition, while a positive slope implies a second-order transition. Figure \ref{Fig 2-1}(b) shows the mean-field behavior of CrTe$_{1.38}$, where the positive slopes near the phase transition confirm a second-order transition. In a magnetic phase transition, the spontaneous magnetization $M_\text{S}$(0,$T$) below $T_\text{C}$, the inverse initial susceptibility $\chi^{-1}_0$(0,$T$) above $T_\text{C}$, and the magnetization isotherm $M(H,T_\text{C})$ at $T_\text{C}$ are characterized by a set of critical exponents $\beta$, $\gamma$, and $\delta$ \cite{Stanley1971, Fisher1967}.

These critical exponents are derived from a series of functions involving the spontaneous magnetization \(M_{S}\) and the initial susceptibility \(\chi_{0}\) near the magnetic 
ordering temperature,\cite{PhysRevLett.19.786, Fisher1967, RevModPhys.71.S358}
\begin{equation}
\begin{split}
    M_\text{S}(T) = M_{0}(-\epsilon)^{\beta}, \epsilon < 0, T<T_\text{C}\\
    \chi^{-1}_{0}(T) = (\frac{h_{0}}{M_{0}})\epsilon^{\gamma}, \epsilon>0, T>T_\text{C}\\
    M=DH^{\frac{1}{\delta}}, \epsilon=0, T=T_\text{C}
\label{equ5}
\end{split}
\end{equation}

where $\epsilon=\frac{(T-T_\text{C})}{T_\text{C}}$ is the reduced temperature, $\frac{h_{0}}{M_{0}}$ and $D$ are critical amplitudes. 

The Arrott-Noakes equation of state in the asymptotic region is used to determine the critical exponents near the transition temperature \cite{PhysRevLett.19.786}
\begin{equation}
    (H/M)^{\frac{1}{\gamma}} = (T-T_\text{C})/T_\text{C} + (M/M_{1})^{\frac{1}{\beta}}
\label{equ6}
\end{equation}
where $M_1$ is a constant. The 3D Heisenberg model $(\beta = 0.365, \gamma = 1.386)$, the 3D Ising model $(\beta = 0.325, \gamma = 1.24)$, the 3D XY model $(\beta = 0.345, \gamma = 1.316)$, and the tricritical mean-field model $(\beta = 0.25, \gamma = 1.0)$ were used to construct the modified Arrott plots (MAP), as shown in Fig.~\ref{Fig 2-1}(c-f). The Arrott plots exhibit quasi-straight lines in the high-field region, whereas a small deviation from linearity is observed in the low-field region as the magnetic domains are not unidirectional \cite{PhysRevB.104.094405}. $M_\text{S}$($T$) and $\chi^{-1}_{0}$($T$) are obtained from the intercepts with the axes $M^{1/\beta}$ and $(H/M)^{1/\gamma}$, respectively, by extrapolating the high-field region of modified Arrott plots.

\begin{figure}
\centering
\includegraphics[width=1\columnwidth]{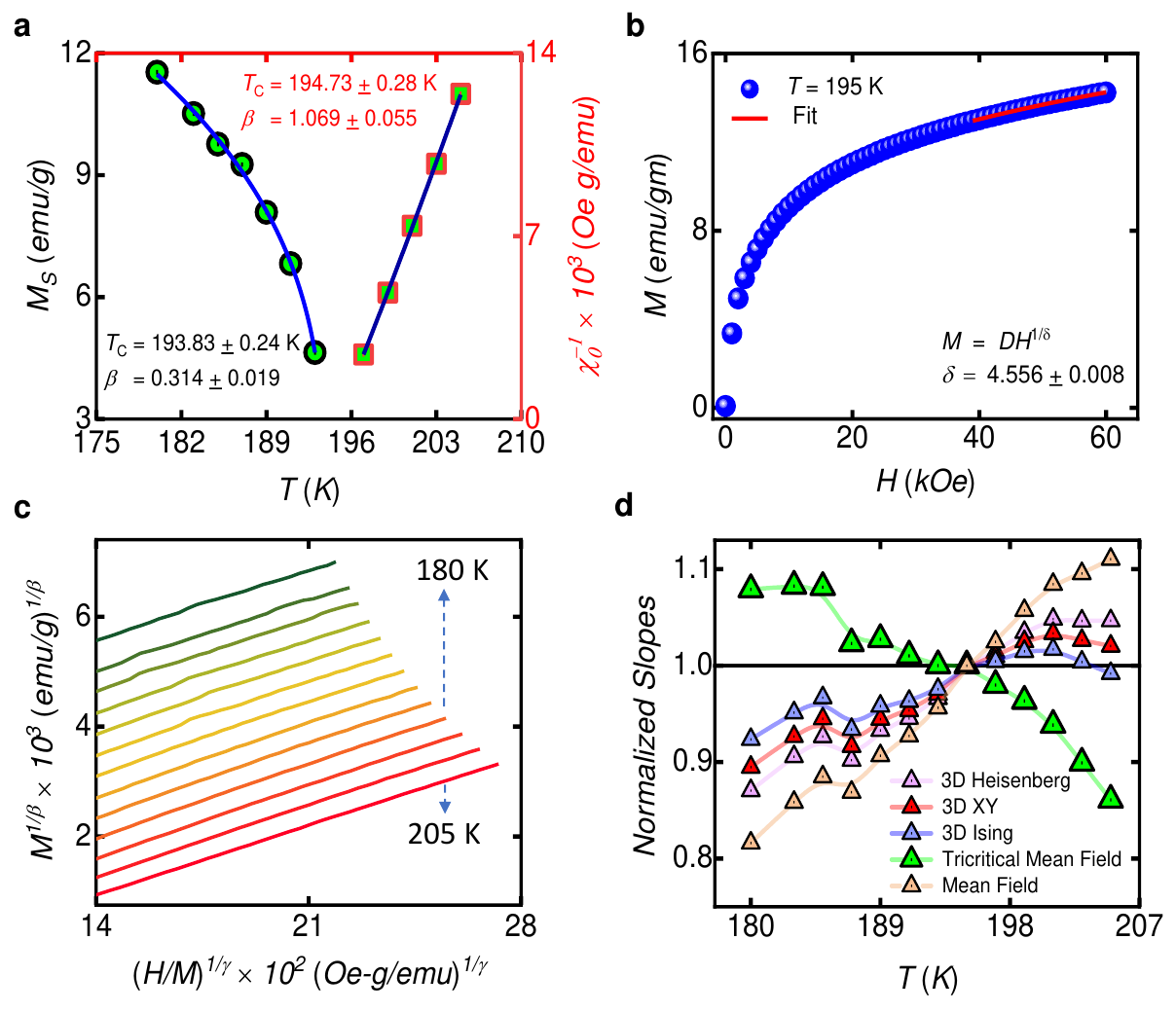}
\caption{ \textbf{Scaling analysis :} (a) Temperature-dependent $($left$)$ the spontaneous magnetization $M_\text{S}$ and $($right$)$ $\chi_{0}^{-1}$ with solid fits based on eq$^n$ \ref{equ5}. (b) $M(H)$ curve collected at $T$ = 195 K and the line (red color) is the fitting of eq$^n$ \ref{equ5}. The critical exponent mentioned in the graph is obtained from fitting. (c). $M^{\frac{1}{\beta}}$ vs $(H/M)^{\frac{1}{\gamma}}$ curves (with MAP) using $\beta$ = 0.314 and $\gamma$ = 1.069. (d). Temperature-dependent normalized slopes (NS) = ${S(T)}$/$S$($T_\text{C}$) for different theoretical models.}
\vspace{-0.45cm}
\label{Fig 3-1}
\end{figure}

To obtain a new set of values $\beta$ and $\gamma$, a rigorous iterative method was adopted \cite{PhysRevB.79.214426}. This yields the values of $\beta$ $\sim$ 0.314, $\gamma$ $\sim$ 1.069, and $T_\text{C}$ $\sim$ 194 K from the final modified Arrott plot [See Fig \ref{Fig 3-1}(a)]. Another critical exponent $\delta$ ($\sim$~4.556) can be determined from the linear fitting of the high field region of the initial $M(H)$ curve [See Fig \ref{Fig 3-1}(b)] at the temperature ($T_\text{C}$ $\sim$ 195 K).  The obtained critical exponents should satisfy the Widom scaling law to demonstrate their self-consistency \cite{Widom1965}.
\begin{equation}
   \delta = 1+\frac{\gamma}{\beta}
\end{equation}
The exponent ($\delta$) is recalculated using the above equation, and the values of $\beta$ and $\gamma$ are obtained from the modified Arrott plot analysis. The $\delta$ $\sim$ 4.404 is closely matched with the value obtained from the critical isotherm analysis, indicating self-consistency and reliability.

For 2D magnets, the $\beta$ typically falls within the range of 0.1 $\leq\beta\leq$ 0.25, according to a comprehensive study of critical exponents by Taroni et al \cite{Taroni2008}. The obtained $\beta$ value for our system suggests a 3D critical behavior. Materials such as Cr$_{2}$Si$_{2}$Te$_{6}$, Cr$_{2}$Ge$_{2}$Te$_{6}$ have $\beta$ values within this window, exhibit 2D Ising type coupling \cite{Liu2016, PhysRevB.96.054406}. However, for systems like Cr$_{1+x}$Te$_{2}$, Cr$_{3}$Te$_{4}$ systems, $\beta$ value is found to be $\geq$ 0.25, indicating a 3D critical phenomenon \cite{PhysRevB.110.L060405, PhysRevB.108.094429}. Subsequently, we constructed the modified Arrott plot [see Fig.~\ref{Fig 3-1}(c)] using the obtained $\beta$ and $\gamma$ values in the high-field region. The $M^{1/\beta}$ vs $(H/M)^{1/\gamma}$ plot in the high-field region clearly displays parallel straight lines for $T$ < $T_\text{C}$ and $T$ > $T_\text{C}$, confirming the reliability of the obtained critical exponents.

\begin{table*}
\caption{\label{tab:table2}  Comparison of the critical exponents of CrTe$_{1.38}$ with  
MAP (modified Arrott plot) approach} 
\begin{ruledtabular}
\begin{tabular}{ccccccc}
 Composition&Technique&Reference&$T_\text{C}(K)$ &$\beta$&$\gamma$&$\delta$\\ \hline
 Cr$_{x}$Te$_{1.38}$&MAP&This work&194.0&0.314&1.069&4.556$^{cal}$\\
 Cr$_{0.62}$Te&MAP&\cite{PhysRevB.96.134410}&230.76&0.314&1.83&6.83\\

 Cr$_{4}$Te$_{5}$&MAP&\cite{PhysRevB.101.214413}&319.06&0.388&1.29&4.32\\

3D-Heisenberg&Theory&\cite{Kaul1985}&--&0.365&1.386&4.8\\
3D-XY&Theory&\cite{Kaul1985}&--&0.345&1.316&4.81\\
3D-Ising&Theory&\cite{Kaul1985}&--&0.325&1.24&4.82\\
Mean-field&Theory&\cite{Kaul1985}&--&0.5&1.0&3.0\\
Tricritical mean field&Theory&\cite{huang2008statistical}&--&0.25&1.0&5.0
\end{tabular}
\end{ruledtabular}
\end{table*}

However, determining the most appropriate model to describe the critical behavior of a system can be challenging. For an ideal model, the modified Arrott plot should display a series of parallel lines in the high-field region, with the slope corresponding to that of the sample. To quantify this, we calculate the normalized slope (NS) = $\frac{S(T)}{S(T_\text{C})}$, which serves as an effective criterion for identifying the best model to describe the system's critical behavior [see Fig.~\ref{Fig 3-1}(d)]. For our as studied system, the NS value of the 3D Ising model is closest to 1, compared to other theoretical models.

\begin{figure}
\centering
\includegraphics[width=1\columnwidth]{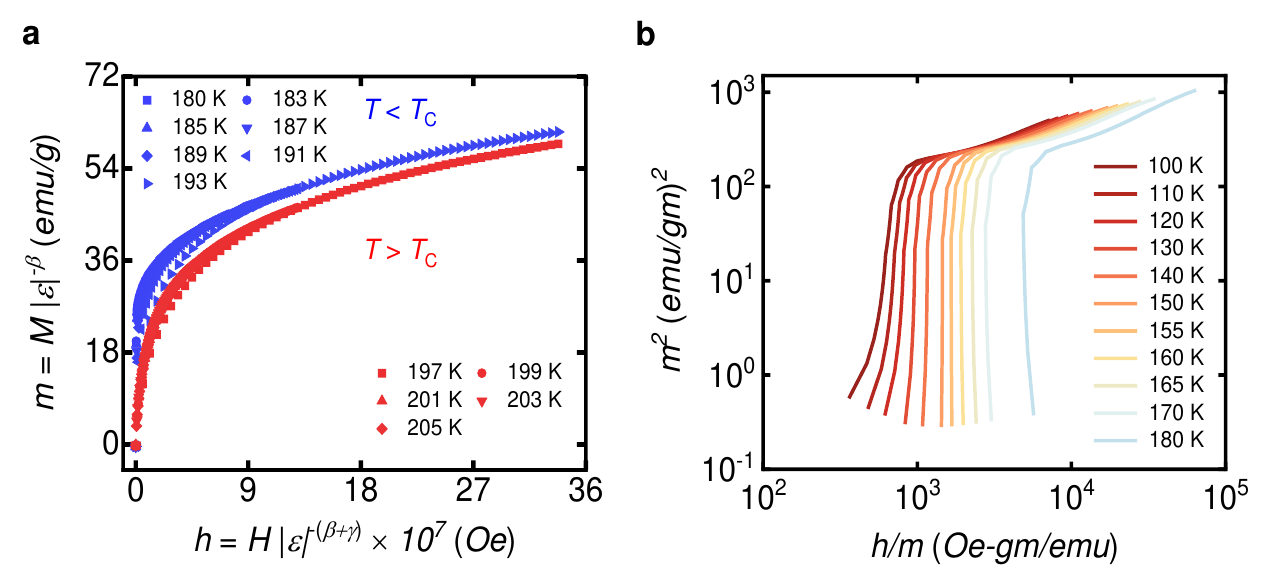}
\caption{\textbf{Scaled plot :} (a) The scaled plot of $m$ versus $h$ around $T_\text{C}$ at specified temperatures. (b) $m^{2}$ vs $h/m$ in log-log scale for the low-field region below transition temperature.}
\vspace{-0.45cm}
\label{Fig 4-1}
\end{figure}

The universality class of the magnetic phase transition can be determined by the exchange coupling $J(r)$. Ghosh et al. \cite{PhysRevLett.81.4740} and Fisher et al. \cite{PhysRevLett.29.917} theoretically modeled magnetic ordering as an attractive interaction between spins. They employed a renormalization group theory analysis, which resulted in an expression for the exchange coupling $J(r)$  that depends on the spatial distance $r$.
\begin{equation}
    J(r)\approx r^{-(d+\sigma)}
\label{equ9}
\end{equation}
where $d$ is the spatial dimensionality and $\sigma$ is a critical exponent which can be calculated as \cite{PhysRevB.65.064443, PhysRevLett.29.917}
\begin{equation}
\begin{split}
    \gamma=1+\frac{4}{d}\frac{n+2}{n+8}\Delta \sigma + \frac{8(n+2)(n-4)}{d^{2}(n+8)^{2}} \\
    \times [1+\frac{2G(\frac{d}{2})(7n+20)}{(n-4)(n+8)}]\Delta \sigma^{2}
\end{split}
\end{equation}
where $\Delta \sigma$ = ($\sigma$-$\frac{d}{2}$), G($\frac{d}{2}$) = 3 - $\frac{1}{4}$ $(\frac{d}{2})^{2}$ and $n$ is the spin dimensionality. For our case, we need to calculate the lattice dimensionality ($d$) and spin dimensionality $n$. To determine the values of $d$ and $n$, we adopt a procedure similar to that outlined here \cite{PhysRevB.65.064443}.

We have chosen the values of {${d:n}$} in such a way that they yield the experimentally observed value of $\gamma$ = 1.069. The procedure was carried out with different sets of ${d:n}$, and we ultimately obtained a value of $\gamma$ close to 1.069 with the combination of ${d:n}={3:1}$ and $\sigma$= 1.651. For $\sigma$ = 1.651, the magnetic interaction decays with spatial distance as $J$($r$) $\approx$ $r^{-4.651}$, indicating a long-range interaction. The nature of the spin interactions is long-range for $\sigma$ < 2 and short-range for $\sigma$ > 2 respectively \cite{PhysRevLett.29.917}. To check the reliability and accuracy of the obtained critical exponents, it is essential to determine whether the exponents yield a magnetic equation of state in the asymptotic critical region. We define two parameters: (i) renormalized magnetization $m$ $\equiv$ $\epsilon^{-\beta}M(H,\epsilon)$, (ii) renormalized field $h$ $\equiv$ $H$$\epsilon^{-(\beta+\gamma)}$. The scaling relation can then be expressed as \cite{Stanley1971}:
\begin{equation}
    M(H,\epsilon)=\epsilon^{\beta} f_{\pm}(H/\epsilon^{\beta+\gamma})
\label{equ9}
\end{equation}
where f$_+$ for $T$ > $T_\text{C}$ and f$_-$ for $T$ < $T_\text{C}$ represent regular functions. The above scaling equation can be reduced to $m$ = $f_{\pm}(h)$ by defining rescaled magnetization $m$ = $M$|$\epsilon|^{-\beta}$ and rescaled field $h$ = $H$|$\epsilon|^{-(\beta+\gamma)}$. According to the scaling relation, the appropriate choice of critical exponents $\beta$, $\gamma$ yield two universal branches for $T$ < $T_\text{C}$ and $T$ > $T_\text{C}$ \cite{PhysRevB.65.064443}. The isothermal magnetization curves around the transition temperature are replotted and shown in Fig \ref{Fig 4-1}(a). All the curves form two branches in the high-field region, corresponding to  $T$<$T_\text{C}$ and $T$>$T_\text{C}$. Fig. \ref{Fig 4-1}(b) clearly exhibits a series of vertical lines in the lower field region. The observed deviations could be attributed to (i) a field-induced phase transition and (ii) the presence of Dzyaloshinskii–Moriya interaction (DMI) at low fields \cite{PhysRevB.91.024403, Dai_2019}. This provides additional evidence for the existence of Néel-type skyrmion bubbles in the temperature range of 100 K to 170 K within this system \cite{Pradhan2024}.

\section{Conclusion}
In summary, we have conducted a comprehensive study of the critical region near 200 K in the itinerant ferromagnet CrTe$_{1.38}$. A clear transition from the paramagnetic
to ferromagnetic phase at $T_\text{C}$ $\approx$ 194 K has been observed. This magnetic transition is also reflected in the temperature-dependent resistivity measurements.
From the Arrott plot, this transition is identified as second-order in nature.  The critical exponents obtained from the analysis of critical behavior near the PM-FM transition are: 
$\beta$ = 0.314, $\gamma$ = 1.069, and $\delta$ = 4.556.. The value of $\beta$ is close to that of the 3D Ising model, while $\gamma$ lies between the mean-field and 3D Ising models. Consequently, these exponents do not correspond to any conventional universality class of homogeneous ferromagnets, although they satisfy the scaling relations and equations for magnetic states predicted by the scaling hypothesis. Furthermore, the obtained 
$J$($r$) $\approx$ $r^{-4.651}$ suggests long-range magnetic interactions. Additionally, a series of vertical lines in the lower-field region of the initial magnetization curves below further $T_\text{C}$ supports the presence of a skyrmion phase in this compound. 
Therefore, our study provides valuable insights into the complex magnetic interactions in the skyrmion-hosting material CrTe$_{1.38}$ through critical scaling analysis.\\

\textbf{Acknowledgment:} This work is supported by the National Key R\&D Program of China (grant no. 2022YFA1203902), the National Natural Science Foundation of China (NSFC) (grant nos. 12374108 and 12104052, 52373226), the Fundamental Research Funds for the Central Universities, and the State Key Lab of Luminescent Materials and Devices, South China University of Technology. Tuhin Debnath thanks UGC, India for research fellowship.

\textbf{Conflict of Interest:}
The authors declare that they have no conflict of interest.\\

\textbf{Data availability:} The data supporting the findings of this study are available from the corresponding author upon
reasonable request.

%

\end{document}